\documentclass[10pt,a4paper,twocolumn,superscriptaddress,accepted=2018-08-02]
  {quantumarticle}

\pdfoutput=1

\usepackage[UKenglish]{babel}
\usepackage{amsmath}
\usepackage{amssymb}
\usepackage[retainorgcmds]{IEEEtrantools}
\usepackage[bb=boondox]{mathalfa}
\usepackage{microtype}
\usepackage[square,comma,numbers,sort&compress]{natbib}
\usepackage{quant_defs}

\def\amjphys{Am.\ J.\ Phys.}%
\def\jphysa{J. Phys.\ A: Math.\ Theor.}%
\def\jphysagen{J.\ Phys.\ A: Math.\ Gen.}
\def\nature{Nature}%
\def\natcomm{Nat.\ Commun.}%
\def\njp{New J.\ Phys.}%
\def\peerjcs{PeerJ Computer Science}%
\def\physics{Physics}%
\def\pra{Phys.\ Rev.\ A}%
\def\prd{Phys.\ Rev.\ D}%
\def\prl{Phys.\ Rev.\ Lett.}%
\def\prx{Phys.\ Rev.\ X}%
\def\rmp{Rev.\ Mod.\ Phys.}%

\DeclareMathOperator{\sign}{sign}
\DeclareMathOperator{\re}{Re}

\newcommand{\rA}{\mathrm{A}}
\newcommand{\rB}{\mathrm{B}}
\newcommand{\rC}{\mathrm{C}}
\newcommand{\rE}{\mathrm{E}}
\newcommand{\rAB}{\mathrm{AB}}
\newcommand{\rAC}{\mathrm{AC}}
\newcommand{\rBC}{\mathrm{BC}}
\newcommand{\rAE}{\mathrm{AE}}
\newcommand{\rABC}{\mathrm{ABC}}
\newcommand{\rABE}{\mathrm{ABE}}
\newcommand{\rABCE}{\mathrm{ABCE}}
\newcommand{\guess}{\text{g}}
\newcommand{\merminr}{M}
\newcommand{\mermini}{M'}
\newcommand{\svetlichny}{M_{+}}
\newcommand{\hilb}{\mathcal{H}}
\newcommand{\vect}[1]{\boldsymbol{#1}}

\title{Randomness versus nonlocality in the Mermin-Bell \newline
  experiment with three parties}
\author[1]{Erik Woodhead}
\orcid{0000-0001-7696-4461}
\email{Erik.Woodhead@icfo.eu}
\author[1]{Boris Bourdoncle}
\orcid{0000-0002-2686-833X}
\author[1,2]{Antonio Ac\'\i{}n}
\orcid{0000-0002-1355-3435}
\affil[1]{ICFO -- Institut de Ci\`e{}ncies Fot\`o{}niques, The Barcelona
  Institute of Science and Technology,
  08860 Castelldefels (Barcelona), Spain}
\affil[2]{ICREA -- Instituci\'o{} Catalana de Recerca i Estudis Avan\c{c}ats,
  Lluis Companys 23, 08010 Barcelona, Spain}
\date{6~August~2018}

\begin{document}

\begin{abstract}
  The detection of nonlocal correlations in a Bell experiment implies almost
  by definition some intrinsic randomness in the measurement outcomes. For
  given correlations, or for a given Bell violation, the amount of randomness
  predicted by quantum physics, quantified by the guessing probability, can
  generally be bounded numerically. However, currently only a few exact
  analytic solutions are known for violations of the bipartite
  Clauser-Horne-Shimony-Holt Bell inequality. Here, we study the randomness
  in a Bell experiment where three parties test the tripartite Mermin-Bell
  inequality. We give tight upper bounds on the guessing probabilities
  associated with one and two of the parties' measurement outcomes as a
  function of the Mermin inequality violation. Finally, we discuss the
  possibility of device-independent secret sharing based on the Mermin
  inequality and argue that the idea seems unlikely to work.
\end{abstract}

\maketitle

\section{Introduction}

Quantum theory predicts correlations incompatible with any local
deterministic model \cite{ref:b1964,ref:bc2014}. The detection of nonlocal
correlations in a Bell experiment thus implies at least some randomness in
the measurement outcomes, regardless of the exact physical mechanism by which
the correlations are produced, provided that communication between the sites
is prohibited. Aside from fundamental interest, this observation is the basis
for device-independent quantum cryptography protocols, such as
device-independent quantum key distribution \cite{ref:my1998,ref:ab2007} and
randomness generation \cite{ref:c2006,ref:pam2010,ref:ck2011}, in which the
detection of nonlocal correlations is a necessary condition in order to
guarantee the randomness of the generated key bits or random numbers from the
perspective of an adversary.

The simplest measure of randomness and typically the easiest to bound is the
guessing probability. This is defined as the probability that an additional
observer, who may have partial access to the underlying quantum state, can
correctly anticipate a given measurement outcome or given set of outcomes in
advance. Aside from its direct operational meaning, the guessing probability
is a useful quantity in the analysis of device-independent cryptography
protocols: security proofs of device-independent protocols frequently depend
on a lower bound on the min-entropy (a function of the guessing probability)
or the conditional von Neumann entropy (which the min-entropy is a lower
bound for)
\cite{ref:rgk2005,ref:r2005,ref:mpa2011,ref:pm2013,ref:pm2013b,ref:afd2018}.
In the practically most relevant case where the measurements are made on a
quantum system, a \emph{numeric} method for deriving an upper bound on the
guessing probability exists, based on a hierarchy of relaxations of the
optimisation problem to semidefinite programming problems
\cite{ref:npa2007,ref:nsps2014,ref:bss2014}, for which reliable optimisation
algorithms exist.

Since the determination of guessing-probability bounds by numerical means is
essentially a solved problem, our interest here is in cases where it is
possible to establish a tight analytic bound. Currently, only a few tight
bounds on the guessing probability are known for the
Clauser-Horne-Shimony-Holt (CHSH) \cite{ref:ch1969} inequality. In
\cite{ref:pam2010}, it was shown that an adversary (``Eve'')'s probability of
guessing one of one party's (e.g., ``Alice'''s) measurement outcomes is
bounded by
\begin{equation}
  \label{eq:pa1_chsh}
  P_{\guess}(A_{1} | \rE) \leq \frac{1}{2}
  + \frac{1}{2} \sqrt{2 - S^{2} / 4}
\end{equation}
in terms of the CHSH expectation value
\begin{equation}
  \label{eq:chsh_exp}
  S = \avg{A_{1} B_{1}} + \avg{A_{1} B_{2}} + \avg{A_{2} B_{1}}
  - \avg{A_{2} B_{2}}
\end{equation}
for $\pm 1$-valued measurements $A_{x}$ and $B_{y}$. More recently, Kaniewski
and Wehner \cite{ref:kw2016} have derived the tight upper bound
\begin{equation}
  \label{eq:pguess_kw}
  P_{\guess}(A | \rB) \leq \frac{1}{2}
  + \frac{1}{4} \Bigro{S/2 + \sqrt{2 - S^{2} / 4}}
\end{equation}
on an average probability
$P_{\guess}(A | \rB) = \bigro{P(A_{1} = B_{3}) + P(A_{2} = B_{3})} / 2$ that
the second party (``Bob'') is able to guess Alice's measurement outcome
without knowing which measurement Alice performed, assuming they are chosen
equiprobably.

Beyond the CHSH scenario, guessing-probability bounds have been determined
for violations of bipartite and multipartite chained Bell inequalities
\cite{ref:bkp2006,ref:ag2012}; however these are derived assuming only the
no-signalling constraints and they are not generally tight assuming the
scenario is restricted to correlations and attacks allowed by quantum
physics.

Here, we study the amount of randomness that can be certified in a Bell
experiment with three parties showing a violation of Mermin's tripartite Bell
inequality \cite{ref:m1990}. We report tight bounds for the following two
cases:
\begin{itemize}
  \item The guessing probability $P_{\guess}(A_{1} | \rE)$ associated with
  the measurement outcome at one site, in terms of two independent Mermin
  expectation values.
  \item The guessing probability $P_{\guess}(A_{1} B_{1} | \rE)$ associated
  with measurement outcomes at two sites, for a given violation of one Mermin
  inequality.
\end{itemize}
The results, the inequalities \eqref{eq:pa1_mermin} and
\eqref{eq:pa1b1_mermin}, can be found in section~\ref{sec:results}, following
a summary of the tripartite scenario we consider. We give the proofs of these
in section~\ref{sec:tangents} in the form of sum-of-squares decompositions
for families of Bell operators which correspond to tangents of the bounds,
together with quantum strategies that show that they are attainable. We found
the sum-of-squares decompositions following an approach similar to
\cite{ref:bp2015}. We have also included the no-signalling bounds in
appendix~\ref{sec:nosignalling_bounds}. Finally, in
section~\ref{sec:secret_sharing} we discuss the prospect of
device-independent secret-sharing \cite{ref:hbb1999}, mainly pointing out
that the idea seems unlikely to work due to the presence of untrusted parties
participating in the Bell test.

\section{Scenario and results}
\label{sec:results}

Our results apply to the following adversarial Bell scenario: three
cooperating parties, Alice, Bob, and Charlie, and an eavesdropper, Eve, share
a quantum state $\rho_{\rABCE}$ on some Hilbert space
$\hilb_{\rA} \otimes \hilb_{\rB} \otimes \hilb_{\rC} \otimes
\hilb_{\rE}$. Alice, Bob, and Charlie may each perform one of two
measurements indexed $x, y, z \in \{1, 2\}$ on their part of the state, which
yield respective outcomes $a, b, c \in \{+, -\}$. Eve performs a measurement
yielding an outcome $e$, intended to be correlated with one or more of
Alice's, Bob's, and Charlie's outcomes. Generally, we will assume, without
loss of generality, that Eve's measurement has the same number of outcomes as
the number of possible different results that the cooperating parties may
obtain that she wishes to distinguish between. The joint correlations are
summarised by a table of conditional probabilities
\begin{IEEEeqnarray}{l}
  P(abce | xyz) \IEEEnonumber \\
  \qquad = \Tr\bigsq{
    \bigro{\Pi^{\rA}_{a|x} \otimes \Pi^{\rB}_{b|y} \otimes \Pi^{\rC}_{c|z}
      \otimes \Pi^{\rE}_{e}}
    \rho_{\rABCE}} \,, \IEEEeqnarraynumspace
\end{IEEEeqnarray}
where $\Pi^{\rA}_{a|x}$ is the measurement operator associated with the
outcome $a$ when Alice performs the measurement $x$, and similarly for Bob's,
Charlie's, and Eve's measurement operators $\Pi^{\rB}_{b|y}$,
$\Pi^{\rC}_{c|z}$, and $\Pi^{\rE}_{e}$. The measurements can be assumed to be
projective, since we do not assume any limit on the dimension of the
underlying Hilbert space. The state and measurements are all treated as
unknown except possibly to Eve.

Eve's goal in this setting is to be able to guess one or more of Alice's,
Bob's and/or Charlie's measurement outcomes. The simplest measure of her
ability to do so, the guessing probability, is simply the probability that
Eve's guess is correct. In the simplest case where Eve aims to guess (say)
Alice's $x = 1$ measurement outcome, the (``local'') guessing probability is
the probability that Eve's measurement outcome is the same as Alice's,
\begin{equation}
  \label{eq:pguess1_def}
  P_{\guess}(A_{1} | \rE) = \sum_{a} P_{\rAE}(aa | x=1) \,,
\end{equation}
where $P_{\rAE}(ae|x) = \sum_{bc} P(abce|xyz)$. Other guessing probabilities
are straightforward variations of this. For instance, the guessing
probability associated with Alice's and Bob's joint outcomes for measurements
$x, y = 1$ is
\begin{equation}
  \label{eq:pguess2_def}
  P_{\guess}(A_{1}B_{1}|\rE)
  = \sum_{a, b} P_{\rABE} \bigro{a b (ab)|x = y = 1} \,,
\end{equation}
where we label Eve's (four) possible measurement outcomes $({+}{+})$,
$({+-})$, $({-}{+})$, and $({-}{-})$. Alice, Bob, and Charlie wish to certify
that Eve's ability to guess outcomes is limited (in mathematical terms, that
guessing probabilities like \eqref{eq:pguess1_def} and \eqref{eq:pguess2_def}
must be less than one) using only the information available to them,
encapsulated by the marginal distribution
$P_{\rABC}(abc|xyz) = \sum_{e} P(abce|xyz)$. A necessary but not necessarily
sufficient condition for this is that this marginal distribution does not
admit a local hidden variable model, i.e., it does not admit a factorisation
of the form
\begin{IEEEeqnarray}{l}
  P_{\rABC}(abc | xyz) \IEEEnonumber \\
  \qquad = \sum_{\lambda} p_{\lambda}
  P_{\rA}(a | x; \lambda) P_{\rB}(b | y; \lambda) P_{\rC}(c | z; \lambda) \,,
  \IEEEeqnarraynumspace
\end{IEEEeqnarray}
which is detected if the marginal distribution $P_{\rABC}$ violates a
Bell inequality.

Here, we study the amount of randomness that can be certified in this
tripartite scenario if a violation of the Mermin-Bell inequality is
observed. The Mermin inequality \cite{ref:m1990} $\merminr \leq 2$ holds for
local-hidden-variable models, where the Mermin correlator is
\begin{IEEEeqnarray}{rCl}
  \label{eq:mermin_real}
  \merminr &=& \avg{A_{1} B_{1} C_{1}} - \avg{A_{1} B_{2} C_{2}}
  \IEEEnonumber \\
  &&-\> \avg{A_{2} B_{1} C_{2}} - \avg{A_{2} B_{2} C_{1}} \,,
\end{IEEEeqnarray}
and in turn $\avg{O}$ denotes the expectation value of the observable
quantity $O$. In the quantum case, $\avg{O} = \Tr[O \rho_{\rABC}]$ is given
by the expectation value in the underlying marginal state $\rho_{\rABC}$ and
the dichotomic operators $- \id \leq A_{x}, B_{y}, C_{z} \leq \id$ are
related to the measurement operators by
\begin{IEEEeqnarray}{rCl+rCl}
  A_{x} &=& \Pi^{\rA}_{+|x} - \Pi^{\rA}_{-|x} \,, &
  B_{y} &=& \Pi^{\rB}_{+|y} - \Pi^{\rB}_{-|y} \,, \IEEEnonumber \\
  C_{z} &=& \Pi^{\rC}_{+|z} - \Pi^{\rC}_{-|z} \,. &&&
\end{IEEEeqnarray}
The Mermin inequality is best known for its association with the
Greenberger-Horne-Zeilinger (GHZ) paradox \cite{ref:gh1990}. The maximal
quantum (and algebraic) violation $\merminr = 4$ is attained by measuring
$A_{1} = B_{1} = C_{1} = \sx$ and $A_{2} = B_{2} = C_{2} = \sy$ on the GHZ
state $\ket{\Psi} = (\ket{111} + \ket{222}) / \sqrt{2}$. Violations greater
than $2 \sqrt{2}$ require entanglement between all three sites
\cite{ref:bg2011}.

The Mermin expression $\merminr$ can be obtained as the real part of the
quantity
\begin{equation}
  \bavg{(A_{1} + i A_{2}) (B_{1} + i B_{2}) (C_{1} + i C_{2})} \,.  
\end{equation}
The imaginary part is also a Mermin expression,
\begin{IEEEeqnarray}{rCl}
  \label{eq:mermin_imag}
  \mermini &=& \avg{A_{1} B_{1} C_{2}} + \avg{A_{1} B_{2} C_{1}}
  \IEEEnonumber \\
  &&+\> \avg{A_{2} B_{2} C_{1}} - \avg{A_{2} B_{2} C_{2}} \,,
\end{IEEEeqnarray}
equivalent to \eqref{eq:mermin_real} up to relabelling some of the inputs and
outputs. The sum $\svetlichny = \merminr + \mermini$ is the correlator
appearing in Svetlichny's inequality \cite{ref:s1987}, which was constructed
to always require nonlocality (and thus entanglement) between all three
parties in order to violate.

\begin{figure}[t]
  \centering
  \includegraphics{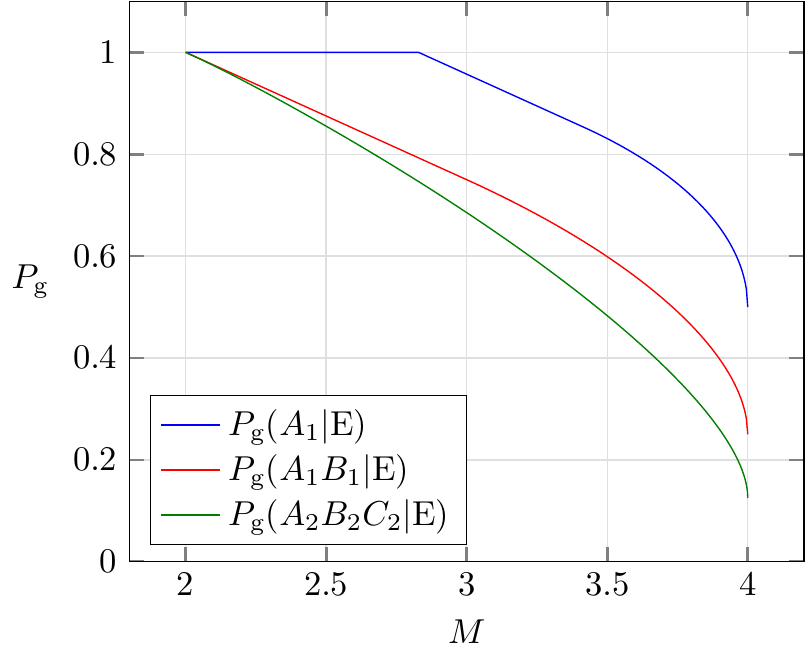}
  \caption{Upper bounds on the guessing probabilities
    $P_{\guess}(A_{1} | \rE)$, $P_{\guess}(A_{1} B_{1} | \rE)$, and
    $P_{\guess}(A_{2} B_{2} C_{2} | \rE)$ for expectation values
    $2 \leq \merminr \leq 4$ of the Mermin expression. The upper bound for
    $P_{\guess}(A_{2} B_{2} C_{2} | \rE)$ was determined numerically at the
    level $1 + \rA^{2} + \rAB + \rAC + \rBC$ of the NPA hierarchy.}
  \label{fig:pguess_mermin}
\end{figure}

\begin{figure}[t]
  \centering
  \includegraphics{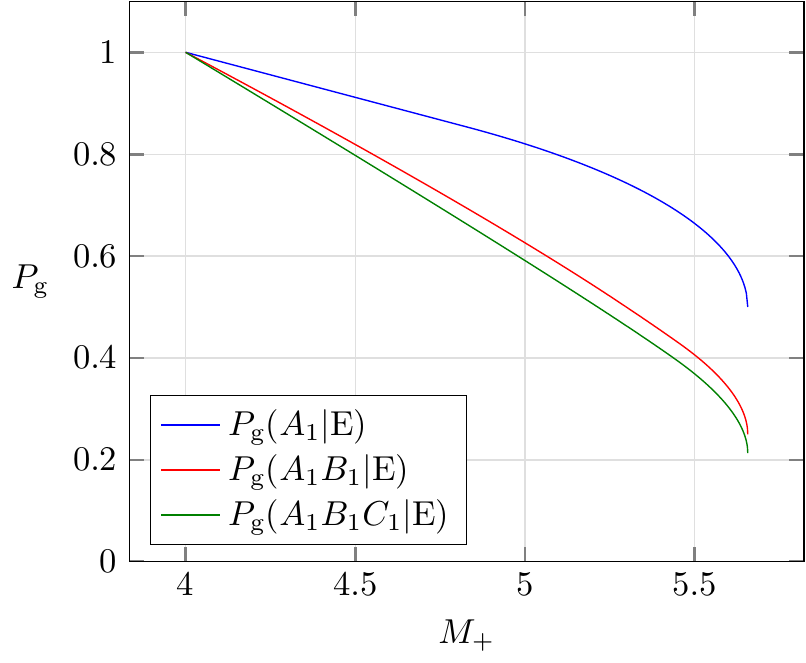}
  \caption{Guessing probabilities for expectation values
    $4 \leq \svetlichny \leq 4 \sqrt{2}$ of the Svetlichny
    expression $\svetlichny = \merminr + \mermini$. The upper bounds 
    for $P(A_{1} B_{1} | \rE)$ and $P(A_{1} B_{1} C_{1} | \rE)$ were obtained
    numerically at levels $1 + \rAB + \rAC + \rBC$ and
    $1 + \rA^{2} + \rAB + \rAC + \rBC$ of the NPA hierarchy.}
  \label{fig:pguess_svetlichny}
\end{figure}

Some randomness bounds, quantified by guessing probabilities involving one,
two, and three parties, are illustrated in figures~\ref{fig:pguess_mermin}
and \ref{fig:pguess_svetlichny} in terms of the Mermin and Svetlichny
expectation values. Of these, we were able to find the analytic form of the
curve for the local guessing probability $P_{\guess}(A_{1} | \rE)$ in both
cases and the curve for $P_{\guess}(A_{1} B_{1} | \rE)$ in terms of the
Mermin expectation value.

For given values of the Mermin or Svetlichny correlators, the corresponding
upper bounds on the local guessing probability have the same functional form,
\begin{equation}
  P_{\guess}(A_{1} | \rE) \leq f(M)
\end{equation}
and
\begin{equation}
  P_{\guess}(A_{1} | \rE) \leq f \bigro{\svetlichny/\sqrt{2}} \,,  
\end{equation}
for the function
\begin{equation}
  f(x) = \begin{cases}
    \frac{1}{2}
    + \frac{1}{2} \sqrt{x (1 - x/4)}
    & \text{if } x \geq 2 + \sqrt{2} \\
    1 + \frac{1}{\sqrt{2}} - x/4
    & \text{if } x \leq 2 + \sqrt{2}
  \end{cases}
\end{equation}
in the range $2 \sqrt{2} \leq x \leq 4$. Both are implied by the tight bound
\begin{equation}
  \label{eq:pa1_mermin}
  P_{\guess}(A_{1} | \rE)
  \leq f \bigro{\sqrt{\merminr^{2} + \mermini^{2}}} \,,
\end{equation}
in which the two Mermin expectation values $\merminr$ and $\mermini$ appear
as independent parameters. Note that since $f$ is a decreasing function in
its argument, \eqref{eq:pa1_mermin} is equivalent to stating that
\begin{equation}
  P_{\guess}(A_{1} | \rE) \leq f \bigro{\cos(\varphi) M + \sin(\varphi) M'}
\end{equation}
holds for all $\varphi$. The result~\eqref{eq:pa1_mermin} certifies some intrinsic
randomness for values of $\merminr$ and $\mermini$ satisfying
\begin{equation}
  2 \sqrt{2} < \sqrt{\merminr^{2} + \mermini^{2}} \leq 4 \,.
\end{equation}
For $\merminr$ alone and the Svetlichny combination
$\svetlichny = \merminr + \mermini$, randomness for one measurement outcome
is certified for $\merminr > 2 \sqrt{2}$ and $\svetlichny > 4$. This is what
one would expect, since these are precisely the ranges that require
entanglement between all three parties to attain. At the boundary
$\sqrt{\merminr^{2} + \mermini^{2}} = 4$, \eqref{eq:pa1_mermin} reduces to
$P_{\guess}(A_{1} | \rE) \leq 1/2$, certifying that the measurement outcome
must be uniformly random.

In the case that the eavesdropper aims to jointly guess two parties'
measurement outcomes, the guessing probability respects the tight bound
\begin{IEEEeqnarray}{l}
  \label{eq:pa1b1_mermin}
  P_{\guess}(A_{1} B_{1} | \rE) \IEEEnonumber \\
  \qquad \leq \begin{cases}
    \frac{3}{4}  - \frac{\merminr}{8}
    + \sqrt{3} \sqrt{\frac{\merminr}{8}
      \bigro{\frac{1}{2} - \frac{\merminr}{8}}}
    & \text{if } \merminr \geq 3 \\
    \frac{3}{2} - \frac{\merminr}{4}
    & \text{if } \merminr \leq 3
  \end{cases} \IEEEeqnarraynumspace
\end{IEEEeqnarray}
in the range $2 \leq \merminr \leq 4$. In this case, we detect some
randomness as soon as the local bound $\merminr \leq 2$ is violated. The
maximum possible violation $\merminr = 4$ implies
$P_{\guess}(A_{1} B_{1} | \rE) \leq 1/4$, corresponding to the maximum
possible randomness.

Beyond this we did not find any new tight bounds for violations of the Mermin
inequality. The upper bound for the global guessing probability
$P(A_{1}B_{1}C_{1}|\rE)$ in terms of $\merminr$ is exactly the same as
\eqref{eq:pa1b1_mermin}, while the upper bound for $P(A_{2}B_{2}C_{2}|\rE)$
(which should attain $1/8$ if the Mermin inequality is maximally violated
\cite{ref:sg2001,ref:ww2001}) appears to be the solution to the maximisation
problem
\begin{IEEEeqnarray}{u+l}
  maximise &
  \begin{IEEEeqnarraybox}[][t]{rl}
    \frac{1}{8} \Bigro{& 1 + 24 \cos \bigro{\tfrac{3}{2} \theta_{2}}
      \alpha \beta \\
      &+\> 2 \cos(3 \theta_{2}) \alpha^{2} + 30 \beta^{2}}
  \end{IEEEeqnarraybox} \IEEEnonumber \\
  subject to & \merminr =
  \begin{IEEEeqnarraybox}[][t]{l}
    \bigro{2 \cos(3 \theta_{1})
      - 6 \cos(\theta_{1} + 2 \theta_{2})} \alpha^{2} \\
    - 12 \cos(\theta_{1} - \theta_{2}) \beta^{2}
  \end{IEEEeqnarraybox} \IEEEeqnarraynumspace \IEEEnonumber \\
  and & 2 \alpha^{2} + 6 \beta^{2} = 1
\end{IEEEeqnarray}
over $\alpha, \beta, \theta_{1}, \theta_{2} \in \mathbb{R}$, which we were
unable to significantly simplify further (let alone
prove). Eq.~\eqref{eq:pa1b1_mermin} also does not generalise in terms of
$\merminr$ and $\mermini$ in the way that the local-guessing-probability
bound does. The upper bound for $P_{\guess}(A_{1} B_{1} | \rE)$ in terms of
the Svetlichny combination (illustrated in
figure~\ref{fig:pguess_svetlichny}) for instance has a different form than
\eqref{eq:pa1b1_mermin}. This is expected since the
local-guessing-probability bound is already less than 1 for any violation of
the local bound, and we were not much more successful in attempting to
identify it analytically than we were for
$P_{\guess}(A_{2} B_{2} C_{2} | \rE)$ in terms of $\merminr$.

For simplicity we have stated the results \eqref{eq:pa1_mermin} and
\eqref{eq:pa1b1_mermin} for the guessing probabilities
$P_{\guess}(A_{1} | \rE)$ and $P_{\guess}(A_{1} B_{1} | \rE)$; however
symmetries of the Mermin correlator(s) imply that the bounds are the same
regardless of what measurements are considered. For the global guessing
probabilities there are two inequivalent cases,
$P_{\guess}(A_{1} B_{1} C_{1} | \rE)$ and
$P_{\guess}(A_{2} B_{2} C_{2} | \rE)$, in terms of $\merminr$.

In figures~\ref{fig:pguess_mermin} and \ref{fig:pguess_svetlichny} we have
also included upper bounds on guessing probabilities for which we do not have
an exact analytic expression. We derived these numerically by solving the
semidefinite programming relaxations at the levels of the
Navascu\'e{}s-Pironio-Ac\'\i{}n (NPA) hierarchy indicated in the figure
captions. We used the arbitrary-precision solver SDPA-GMP
\cite{ref:sdpa,ref:n2010} for this purpose. We have made the code we used to
generate the relaxations available online \cite{ref:npa-hierarchy}.

\section{Tangent Bell expressions}
\label{sec:tangents}

We have asserted that the local and two-party guessing probabilities respect
the upper bounds \eqref{eq:pa1_mermin} and \eqref{eq:pa1b1_mermin} and that
the bounds are tight. We prove these assertions in this section.

\subsection{General idea illustrated with CHSH}

Proving the main results \eqref{eq:pa1_mermin} and \eqref{eq:pa1b1_mermin} is
equivalent to proving families of linear inequalities corresponding to
tangents of the curves. We illustrate the approach using CHSH as an example,
for which this has already been done \cite{ref:mpa2011,ref:amp2012}. It was
shown in \cite{ref:amp2012} that the quantum expectation value of a modified
CHSH expression respects the tight upper bound
\begin{equation}
  \label{eq:ibeta_tsirelson}
  \beta \avg{A_{1}} + S \leq 2 \sqrt{2} \sqrt{1 + \beta^{2}/4}
\end{equation}
in the parameter range $0 \leq \beta \leq 2$, where
$S = \avg{A_{1} B_{1} + A_{1} B_{2} + A_{2} B_{1} - A_{2} B_{2}}$ is the CHSH
expectation value. Eq.~\eqref{eq:ibeta_tsirelson} can be rewritten as an
upper bound
\begin{equation}
  \avg{A_{1}} \leq \frac{1}{\beta} \Bigro{
    2 \sqrt{2} \sqrt{1 + \beta^{2}/4} - S}
\end{equation}
for $\avg{A_{1}}$. Assuming that $S \geq 2$, minimising the right-hand side
over $\beta$ produces the tightest possible bound
\begin{equation}
  \avg{A_{1}} \leq \sqrt{2 - S^{2}/4} \,.
\end{equation}
This bound has two key characteristics. First, since the CHSH expression
remains unchanged under (for example) the replacements $A_{x} \mapsto -A_{x}$
and $B_{y} \mapsto -B_{y}$, the same upper bound holds for $-\avg{A_{1}}$ as
well as $\avg{A_{1}}$. Second, the right-hand side is by construction concave
in $S$, since it is derived by minimising over a family of hyperplanes. Using
these properties and that $P_{\rA}({+} | x) = (1 + \avg{A_{x}})/2$ and
$P_{\rA}({-} | x) = (1 - \avg{A_{x}})/2$, the result is quickly obtained:
\begin{IEEEeqnarray}{rCl}
  \label{eq:pa1_chsh_derivation}
  P_{\guess}(A_{1} | \rE)
  &=& \sum_{a} P_{\rAE}(a a | x=1) \IEEEnonumber \\
  &=& \sum_{a} P_{\rE}(a) P_{\rA|\rE}(a | x=1, e=a) \IEEEnonumber \\
  &\leq& \sum_{a} P_{\rE}(a) \frac{1}{2}
  \Bigro{1 + \sqrt{2 - S\du{|a}{2}/4}} \IEEEnonumber \\
  &\leq& \frac{1}{2} + \frac{1}{2} \sqrt{2 - S^{2} / 4} \,,
\end{IEEEeqnarray}
where $S_{|a}$ in the third line is the CHSH expectation value conditioned on
Eve obtaining the outcome $e = a$.

In passing, we mention that the bound \eqref{eq:pguess_kw} for
$P_{\guess}(A | \rB) = \frac{1}{2} + \frac{1}{4} \avg{(A_{1} + A_{2}) B_{3}}$
can similarly be derived from the inequality
\begin{equation}
  \label{eq:kw_linearisation}
  \alpha \avg{(A_{1} + A_{2}) B_{3}} + S
  \leq 2 \sqrt{1 + (1 + \alpha)^{2}}
\end{equation}
for $\alpha \geq 0$. The inequality \eqref{eq:kw_linearisation} itself is
implied by the tight quantum bound derived for the $I_{\alpha}^{\beta}$
expression in \cite{ref:amp2012}, since there is clearly no advantage for the
operator $B_{3}$ to be different from $B_{1}$ in order to maximise the
left-hand side.

The same general approach works for the main results of
section~\ref{sec:results}. The Mermin expectation values $\merminr$ and
$\mermini$ are both symmetric under the transformations
$A_{x}, C_{z} \mapsto -A_{x}, -C_{z}$ and
$B_{y}, C_{z} \mapsto -B_{y}, -C_{z}$. These can be used to map the
probability $P_{\rA}({+} | 1)$ to $P_{\rA}({-} | 1)$ and the probability
$P_{\rAB}({+}{+} | 11)$ to any of the probabilities $P_{\rAB}({+}{-} | 11)$,
$P_{\rAB}({-}{+} | 11)$, and $P_{\rAB}({-}{-} | 11)$, and vice
versa. Consequently, in order to derive upper bounds on
$P_{\guess}(A_{1} | \rE)$ and $P_{\guess}(A_{1} B_{1} | \rE)$, we need only
derive concave upper bounds for
\begin{equation}
  P_{\rA}({+} | 1) = \frac{1}{2} \bigro{1 + \avg{A_{1}}}
\end{equation}
and
\begin{equation}
  P_{\rAB}({+}{+} | 11)
  = \frac{1}{4} \bigro{1 + \avg{A_{1}}
    + \avg{B_{1}} + \avg{A_{1} B_{1}}} \,.
\end{equation}

\subsection{Local guessing probability linearisation}
\label{sec:pa1_linearisation}

Similarly to the derivation for CHSH summarised above, the
local-guessing-probability bound \eqref{eq:pa1_mermin} for
$\sqrt{\merminr^{2} + \mermini^{2}} \geq 2 \sqrt{2}$ is implied by the
linearisation
\begin{multline}
  \label{eq:a1_mermin_lin}
  \cos(\theta) \avg{A_{1}}
  + \frac{1}{2} \sin(\theta)
  \bigro{\cos(\varphi) \merminr + \sin(\varphi) \mermini} \\
  \leq 1 + \sin(\theta) \,,
\end{multline}
which holds for $\theta$ in the range $\pi / 4 \leq \theta \leq \pi / 2$ and
for all $\varphi$. We can see that \eqref{eq:a1_mermin_lin} is tight by
observing that is attained if (for example) the measurements
\begin{IEEEeqnarray}{c+c}
  \label{eq:measurements_mermin}
  A_{1} = B_{1} = \sx \,, & A_{2} = B_{2} = \sy
  \IEEEeqnarraynumspace
\end{IEEEeqnarray}
and
\begin{IEEEeqnarray}{rCl}
  C_{1} &=& \cos(\varphi) \sx - \sin(\varphi) \sy \,, \\
  C_{2} &=& \sin(\varphi) \sx + \cos(\varphi) \sy
\end{IEEEeqnarray}
are performed on the state
\begin{IEEEeqnarray}{rCl}
  \label{eq:state_a1_mermin}
  \ket{\Psi} &=& \cos(\tfrac{\theta}{2}) \frac{1}{\sqrt{2}}
  \Bigro{\ket{{+}{+}{+}} + \ket{{+}{-}{-}}}
  \IEEEnonumber \\
  &&+\> \sin(\tfrac{\theta}{2}) \frac{1}{\sqrt{2}}
  \Bigro{\ket{{-}{+}{-}} + \ket{{-}{-}{+}}} \,, \IEEEeqnarraynumspace
\end{IEEEeqnarray}
where $\ket{\pm} = (\ket{1} \pm \ket{2}) / \sqrt{2}$ are the eigenstates of
the $\sx$ operator and $\theta$ is the same angle as in
\eqref{eq:a1_mermin_lin}. With this state and measurements, one can readily
verify that $\avg{A_{1}} = \cos(\theta)$ and
$\frac{1}{2} \bigro{\cos(\varphi) \merminr + \sin(\varphi) \mermini}
= 1 + \sin(\theta)$, which attain \eqref{eq:a1_mermin_lin}.

The linearisation \eqref{eq:a1_mermin_lin} ceases to apply for $\theta <
\pi/4$. It is violated, for instance, by measuring
\begin{IEEEeqnarray}{rCl+rCl}
  A_{1} &=& \id \,, & A_{2} &=& -\id \,, \\
  B_{1} &=& \sx \,, & B_{2} &=& \sy \,,
\end{IEEEeqnarray}
and
\begin{IEEEeqnarray}{rCl}
  C_{1} &=& \cos(\varphi) \sx - \sin(\varphi) \sy \,, \\
  C_{2} &=& \sin(\varphi) \sx + \cos(\varphi) \sy
\end{IEEEeqnarray}
on a state $\ket{\Psi'} = \ket{\chi}_{\rA} \ket{\psi}_{\rBC}$, where
$\ket{\chi}$ is any state on Alice's subsystem and Bob and Charlie share the
state
\begin{equation}
  \ket{\psi} = \frac{1}{\sqrt{2}} \Bigro{
    e^{-i\tfrac{\pi}{8}} \ket{11} + e^{i\tfrac{\pi}{8}} \ket{22}} \,.
\end{equation}
This strategy yields $\avg{A_{1}} = 1$ and
\begin{equation}
  \cos(\varphi) \merminr + \sin(\varphi) \mermini
  = 2 \sqrt{2} \,,
\end{equation}
and the right-hand side of \eqref{eq:a1_mermin_lin} attains
$\cos(\theta) + \sqrt{2} \sin(\theta)$. Importantly, for $\theta = \pi/4$, we
see that \eqref{eq:a1_mermin_lin} can be attained with a strategy for which
Alice's $A_{1}$ measurement produces a deterministic outcome.

We prove the linearisation \eqref{eq:a1_mermin_lin} by showing that the
operator
\begin{IEEEeqnarray}{rCl}
  \label{eq:a1_mermin_target}
  T &=& \bigro{1 + \sin(\theta)} \id - \cos(\theta) A_{1} \IEEEnonumber \\
  &&-\> \frac{1}{2} \sin(\theta)
  \bigro{\cos(\varphi) \hat{\merminr} + \sin(\varphi) \hat{\mermini}}
\end{IEEEeqnarray}
is positive semidefinite, where
\begin{IEEEeqnarray}{rCl}
  \hat{\merminr} &=& A_{1} B_{1} C_{1} - A_{1} B_{2} C_{2} \IEEEnonumber \\
  &&-\> A_{2} B_{1} C_{2} - A_{2} B_{2} C_{1} \,, \\
  \hat{\mermini} &=& A_{1} B_{1} C_{2} + A_{1} B_{2} C_{1} \IEEEnonumber \\
  &&+\> A_{2} B_{1} C_{1} - A_{2} B_{2} C_{2} \,. \IEEEeqnarraynumspace
\end{IEEEeqnarray}
A sum-of-squares decomposition that shows this is
\begin{equation}
  \label{eq:a1_mermin_sos}
  T = \abs{P^{+}_{1}}^{2} + \abs{P^{+}_{2}}^{2}
  + \abs{P^{-}_{1}}^{2} + \abs{P^{-}_{2}}^{2}
\end{equation}
where $\abs{O}^{2} = O^{\dagger} O$,
\begin{IEEEeqnarray}{rCl}
  P^{+}_{1} &=& \alpha R^{+}_{1} + \beta R^{+}_{2}
  - \beta R^{+}_{3} - \alpha R^{+}_{4} \,, \\
  P^{+}_{2} &=& \gamma R^{+}_{1} - \delta R^{+}_{3} \,, \\
  \IEEEnonumber \\
  P^{-}_{1} &=& \beta R^{-}_{1} + \alpha R^{-}_{2}
  + \alpha R^{-}_{3} + \beta R^{-}_{4} \,, \\
  P^{-}_{2} &=& \delta R^{-}_{1} + \gamma R^{-}_{3} \,,
\end{IEEEeqnarray}
$R^{\pm}_{i}$ are the operators
\begin{IEEEeqnarray}{rCl}
  \label{eq:a1_Rfirst}
  R^{+}_{1} &=& \cos(\varphi) (B_{1} + C_{1})
  + \sin(\varphi) (B_{2} + C_{2}) \IEEEeqnarraynumspace \IEEEnonumber \\
  &&-\> A_{1} (B_{1} + C_{1}) \,, \\
  R^{+}_{2} &=& \cos(\theta) (B_{2} + C_{2})
  - A_{1} (B_{2} + C_{2}) \IEEEnonumber \\
  &&+\> \sin(\theta) A_{2} (B_{1} + C_{1}) \,, \\
  R^{+}_{3} &=& \sin(\varphi) (B_{1} + C_{1})
  - \cos(\varphi) (B_{2} + C_{2}) \IEEEnonumber \\
  &&-\> A_{1} (B_{2} + C_{2}) \,, \\
  R^{+}_{4} &=& \cos(\theta) (B_{1} + C_{1})
  - A_{1} (B_{1} + C_{1}) \IEEEnonumber \\
  &&-\> \sin(\theta) A_{2} (B_{2} + C_{2}) \,, \\
  \IEEEnonumber \\
  R^{-}_{1} &=& \cos(\varphi) (B_{1} - C_{1})
  + \sin(\varphi) (B_{2} - C_{2}) \IEEEnonumber \\
  &&+\> A_{1} (B_{1} - C_{1}) \,, \\
  R^{-}_{2} &=& \cos(\theta) (B_{2} - C_{2})
  - A_{1} (B_{2} - C_{2}) \IEEEnonumber \\
  &&+\> \sin(\theta) A_{2} (B_{1} - C_{1}) \,, \\
  R^{-}_{3} &=& \sin(\varphi) (B_{1} - C_{1})
  - \cos(\varphi) (B_{2} - C_{2}) \IEEEnonumber \\
  &&+\> A_{1} (B_{2} - C_{2}) \,, \\
  \label{eq:a1_Rlast}
  R^{-}_{4} &=& \cos(\theta) (B_{1} - C_{1})
  - A_{1} (B_{1} - C_{1}) \IEEEnonumber \\
  &&-\> \sin(\theta) A_{2} (B_{2} - C_{2}) \,,
\end{IEEEeqnarray}
and the coefficients $\alpha$, $\beta$, $\gamma$, $\delta$ are related to
$\theta$ and $\varphi$ by
\begin{IEEEeqnarray}{rCrl}
  \alpha &=& \IEEEeqnarraymulticol{2}{l}{
    \frac{\sin \bigro{\tfrac{\varphi}{2}}}{
      4 \cos \bigro{\tfrac{\theta}{2}}} \,,} \\
  \beta &=& \IEEEeqnarraymulticol{2}{l}{
    \frac{\cos \bigro{\tfrac{\varphi}{2}}}{
      4 \cos \bigro{\tfrac{\theta}{2}}} \,,} \\
  \label{eq:gamma_sol}
  \gamma &=& \frac{1}{4} \Bigsq{\sin(\theta)
    + \cos(\theta) \cos(\varphi) & \IEEEnonumber \\
    &&-\> \sin(\varphi) \sqrt{- \cos(2 \theta)}&}^{1/2} \,, \\
  \label{eq:delta_sol}
  \delta &=& \frac{s}{4} \Bigsq{\sin(\theta)
    - \cos(\theta) \cos(\varphi) & \IEEEnonumber \\
    &&+\> \sin(\varphi) \sqrt{- \cos(2 \theta)}&}^{1/2} \,,
\end{IEEEeqnarray}
where $s = \pm 1$ in the last line is the sign
\begin{equation}
  s = - \sign \Bigro{\cos(\theta) \sin(\varphi)
    + \cos(\varphi) \sqrt{-\cos(2 \theta)}} \,.
\end{equation}
The $R^{\pm}_{i}$s have been grouped by whether or not they change sign under
the replacements
\begin{equation}
  \label{eq:a1_mermin_symmetry}
  \tau \colon \left\{
    \begin{IEEEeqnarraybox}[][c]{rCl} 
      B_{1} &\mapsto& C_{1} \\
      B_{2} &\mapsto& C_{2} \\
      C_{1} &\mapsto& B_{1} \\
      C_{2} &\mapsto& B_{2}
    \end{IEEEeqnarraybox}
  \right. \,,
\end{equation}
which is a symmetry of \eqref{eq:a1_mermin_target}.

The parameters $\gamma$ and $\delta$ are chosen to solve the simultaneous
equations
\begin{IEEEeqnarray}{rCl}
  \label{eq:gamma_delta_1}
  8 \gamma^{2} + 8 \delta^{2} - \sin(\theta) &=& 0 \,, \\
  \label{eq:gamma_delta_2}
  8 \cos(\varphi) \gamma^{2}
  - 16 \sin(\varphi) \gamma \delta \IEEEnonumber \\
  -\> 8 \cos(\varphi) \delta^{2} - \cos(\theta) &=& 0 \,,
\end{IEEEeqnarray}
which we encountered when searching for a decomposition. They are solvable
for real-valued $\gamma$ and $\delta$ (and \eqref{eq:gamma_sol} and
\eqref{eq:delta_sol} are solutions) if $\sin(\theta)$ is positive and greater
than $\abs{\cos(\theta)}$, which is the case for the range
$\pi/4 \leq \theta \leq \pi/2$ of values of $\theta$ for which we need to
show that the linearisation \eqref{eq:a1_mermin_lin} holds. It is not
difficult to check in this case that
\begin{equation}
  \sin(\theta) - \abs{\cos(\theta) \cos(\varphi)}
  \geq \abs{\sin(\varphi)} \sqrt{-\cos(2 \theta)}
\end{equation}
holds for arbitrary $\varphi$, verifying that the expressions under the outer
square roots in \eqref{eq:gamma_sol} and \eqref{eq:delta_sol} are
nonnegative. The operators $P^{\pm}_{i}$ are then all Hermitian and
$\abs{P^{\pm}_{i}}^{2}$ can be simplified to $P^{\pm\,2}_{i}$.

The Python script \texttt{pa1\_mermin\_sos.py} supplied with this article
uses the SymPy library \cite{ref:sympy} to verify symbolically that the
sum-of-squares decomposition \eqref{eq:a1_mermin_sos} expands to
\eqref{eq:a1_mermin_target}, under the assumption that the operators
$P^{\pm}_{i}$ are Hermitian and that the conditions \eqref{eq:gamma_delta_1}
and \eqref{eq:gamma_delta_2} for $\gamma$ and $\delta$ can be satisfied.

\subsection{Two-party guessing probability linearisation}
\label{sec:pa1b1_linearisation}

For $\merminr \geq 2$, the guessing-probability bound \eqref{eq:pa1b1_mermin}
follows from the linearisation
\begin{equation}
  \label{eq:a1b1_mermin_lin}
  \beta \avg{A_{1} + B_{1} + A_{1} B_{1}}
  + \alpha \merminr \leq \gamma \,,
\end{equation}
where
\begin{IEEEeqnarray}{rCl}
  \label{eq:beta_lambda_mu}
  \beta &=& (\lambda - \mu) (\lambda + 3 \mu) \,, \\
  \label{eq:alpha_lambda_mu}
  \alpha &=& 4 \lambda \mu \,, \\
  \label{eq:gamma_lambda_mu}
  \gamma &=& (3 \lambda + \mu) (\lambda + 3 \mu) \,,
\end{IEEEeqnarray}
which holds for parameters $\lambda$ and $\mu$ satisfying
\begin{equation}
  3 \mu \geq \lambda \geq \mu \,.
\end{equation}
In the extreme cases $\lambda = 3 \mu$ and $\lambda = \mu$,
\eqref{eq:a1b1_mermin_lin} reduces respectively to
\begin{equation}
  4 P_{\rAB}({+}{+} | 11) + \merminr \leq 6 \,,
\end{equation}
which corresponds to the linear part of \eqref{eq:pa1b1_mermin}, and to the
bound $\merminr \leq 4$ for the Mermin correlator itself, where the gradient
of \eqref{eq:pa1b1_mermin} is infinite. (Eq.~\eqref{eq:a1b1_mermin_lin} also
appears to hold for $0 \leq \lambda < \mu$; however
\eqref{eq:a1b1_mermin_lin} then translates to a lower bound on
$P_{\rAB}({+}{+} | 11)$, which we did not interest ourselves in.)
Eq.~\eqref{eq:a1b1_mermin_lin} is attained with equality by measuring
$A_{1} = B_{1} = C_{1} = \sx$ and $A_{2} = B_{2} = C_{2} = \sy$ on the state
\begin{equation}
  \label{eq:pa1b1_mermin_state}
  \ket{\Psi} = \lambda \ket{{+}{+}{+}}
  + \mu \bigro{\ket{{+}{-}{-}} + \ket{{-}{+}{-}} + \ket{{-}{-}{+}}} \,,
\end{equation}
with $\lambda$ and $\mu$ scaled to satisfy $\lambda^{2} + 3 \mu^{2} = 1$ so
that the state is properly normalised. In this case $P_{\rAB}({+}{+} | 11)$
and $\merminr$ work out to
\begin{equation}
  P_{\rAB}({+}{+} | 11) = \lambda^{2} \,,
\end{equation}
and
\begin{equation}
  \merminr = (\lambda + 3 \mu)^{2} \,;
\end{equation}
these are related by
\begin{equation}
  P_{\rAB}({+}{+} | 11) = \frac{3}{4} - \frac{\merminr}{8}
  + \sqrt{3} \sqrt{\frac{\merminr}{8} \Bigro{
      \frac{1}{2} - \frac{\merminr}{8}}} \,,
\end{equation}
corresponding to the nonlinear part of \eqref{eq:pa1b1_mermin}. The condition
$3 \mu \geq \lambda \geq \mu$ and normalisation $\lambda^{2} + 3 \mu^{2} = 1$
also translate to precisely the ranges $1/4 \leq P({+}{+} | 11) \leq 3/4$ and
$3 \leq \merminr \leq 4$ to which the nonlinear part of
\eqref{eq:pa1b1_mermin} applies.

With the same state \eqref{eq:pa1b1_mermin_state} and optimal measurements,
we also have $P_{\rABC}({+}{+}{+}|111) = \lambda^{2} =
P_{\rAB}({+}{+}|11)$. This implies that the upper bound
\eqref{eq:pa1b1_mermin} for $P_{\guess}(A_{1} B_{1} | \rE)$ is also the tight
upper bound for $P_{\guess}(A_{1} B_{1} C_{1} | \rE)$.

The linearisation \eqref{eq:a1b1_mermin_lin} is equivalent to the operator
inequality
\begin{equation}
  \label{eq:a1b1_mermin_target}
  T = \gamma \id - \beta (A_{1} + B_{1} + A_{1} B_{1})
  - \alpha \hat{M} \geq 0 \,.
\end{equation}
This is shown by the sum-of-squares decomposition
\begin{IEEEeqnarray}{rCl}
  \label{eq:a1b1_mermin_sos}
  T &=& \babs{P^{++}_{1}}^{2} + \babs{P^{++}_{2}}^{2}
  + \babs{P^{++}_{3}}^{2} + \babs{P^{++}_{4}}^{2}
  + \babs{P^{+-}_{2}}^{2} \IEEEnonumber \\
  &&+\> \babs{P^{-+}_{1}}^{2} + \babs{P^{-+}_{2}}^{2} + \babs{P^{--}_{1}}^{2}
  + \babs{P^{--}_{3}}^{2} \,,
\end{IEEEeqnarray}
where
\begin{IEEEeqnarray}{rCl}
  P^{++}_{1} &=& \frac{\sqrt{\lambda + \mu}}{4 \sqrt{\mu}}
  (3 \mu - \lambda) R^{++}_{1} \,, \\
  P^{++}_{2} &=& \frac{1}{4 \sqrt{\mu}}
  \sqrt{(\lambda^{2} - \mu^{2}) (3 \mu - \lambda)} \IEEEnonumber \\
  &&\times\> \bigro{R^{++}_{1} + 2 R^{++}_{2}} \,, \\
  P^{++}_{3} &=& \frac{\sqrt{3 \mu - \lambda}}{
    2 \sqrt{\mu} (\lambda + \mu)} R^{++}_{3} \,, \\
  P^{++}_{4} &=& \frac{1}{2 \sqrt{\lambda \mu}} \Bigro{
    \frac{\lambda - \mu}{\lambda + \mu} R^{++}_{3} + R^{++}_{4}} \,, \\
  \IEEEnonumber \\
  P^{+-}_{2} &=& \frac{1}{2} \sqrt{\lambda (\lambda - \mu)} R^{+-}_{2} \,, \\
  \IEEEnonumber \\
  P^{-+}_{1} &=& \frac{1}{2} \frac{\sqrt{\lambda}}{\sqrt{2 \mu}}
  \sqrt{(\lambda - \mu)^{2} + 4 \mu^{2}} R^{-+}_{1} \,,
  \IEEEeqnarraynumspace \\
  P^{-+}_{2} &=& \frac{1}{2} \sqrt{
    \frac{\lambda (3 \mu - \lambda)}{\mu (\lambda + \mu)}} R^{-+}_{2} \,, \\
  \IEEEnonumber \\
  P^{--}_{1} &=& \sqrt{\frac{\lambda (\lambda - \mu)}{2}} R^{--}_{1} \,, \\
  P^{--}_{3} &=& \frac{
    \sqrt{\lambda (\lambda - \mu)}}{2 (\lambda + \mu)}
  \bigro{R^{--}_{2} + R^{--}_{3}} \,,
\end{IEEEeqnarray}
and
\begin{IEEEeqnarray}{rCl}
  R^{++}_{1} &=& (A_{1} + B_{1}) (\id - C_{1}) \,, \\
  R^{++}_{2} &=& C_{1} - A_{1} B_{1} \,, \\
  R^{++}_{3} &=& (\lambda - \mu)^{2} \id 
  + (\lambda + \mu)^{2} C_{1} \IEEEnonumber \\
  &&-\> (\lambda^{2} - \mu^{2}) (A_{1} + B_{1})
  + 4 \lambda \mu \, A_{2} B_{2} \,, \\
  R^{++}_{4} &=& (\lambda - \mu)^{2} \id
  + \mu (\lambda + \mu) (A_{1} + B_{1})
  \IEEEnonumber \\
  &&-\> (\lambda^{2} - \mu^{2}) C_{1}
  + 2 \lambda \mu \, (A_{2} + B_{2}) C_{2} \,, \\
  \IEEEnonumber \\
  R^{+-}_{1} &=& (\lambda - \mu)^{2} (A_{2} + B_{2})
  - 2 (\lambda - \mu)^{2} C_{2} \IEEEnonumber \\
  &&-\> (\lambda^{2} - \mu^{2}) (A_{2} + B_{2}) C_{1} \IEEEnonumber \\
  &&+\> (\lambda^{2} - \mu^{2}) (A_{1} + B_{1}) C_{2} \,, \\
  R^{+-}_{2} &=& (A_{2} + B_{2}) - 2 C_{2} - (A_{1} B_{2} + A_{2} B_{1})
  \IEEEnonumber \\
  &&+\> (A_{1} + B_{1}) C_{2} \,, \\
  \IEEEnonumber \\
  R^{-+}_{1} &=& (A_{1} - B_{1}) (\id + C_{1}) \,, \\
  R^{-+}_{2} &=& (\lambda + \mu) (A_{1} - B_{1})
  - 2 \mu (A_{2} - B_{2}) C_{2} \,, \IEEEeqnarraynumspace \\
  \IEEEnonumber \\
  R^{--}_{1} &=& (A_{2} - B_{2}) (\id - C_{1}) \,, \\
  R^{--}_{2} &=& 2 \mu (A_{2} - B_{2})
  - (\lambda + \mu) (A_{1} - B_{1}) C_{2} \,, \\
  R^{--}_{3} &=& (\lambda - \mu) (A_{2} - B_{2}) \IEEEnonumber \\
  &&+\> (\lambda + \mu) (A_{1} B_{2} - A_{2} B_{1}) \,.
\end{IEEEeqnarray}
The $R^{\pm \pm'}_{i}$s are grouped according to whether they change sign
under the replacements
\begin{IEEEeqnarray}{c+c}
  \tau_{1} \colon \left\{
    \begin{IEEEeqnarraybox}[][c]{rCl} 
      A_{1} &\mapsto& B_{1} \\
      A_{2} &\mapsto& B_{2} \\
      B_{1} &\mapsto& A_{1} \\
      B_{2} &\mapsto& A_{2}
    \end{IEEEeqnarraybox}
  \right. \,, &
  \tau_{2} \colon \left\{
    \begin{IEEEeqnarraybox}[][c]{rCl} 
      A_{2} &\mapsto& -A_{2} \\
      B_{2} &\mapsto& -B_{2} \\
      C_{2} &\mapsto& -C_{2}
    \end{IEEEeqnarraybox}
  \right. \,. \IEEEeqnarraynumspace
\end{IEEEeqnarray}
Note that we have included an operator, $R^{+-}_{1}$, among the list of
$R^{\pm \pm'}_{i}$s that we attempted to construct a sum-of-squares
decomposition out of, although ultimately we did not use it.

The Python script \texttt{pa1b1\_mermin\_sos.py} checks that the
sum-of-squares decomposition \eqref{eq:a1b1_mermin_sos} expands to
\eqref{eq:a1b1_mermin_target}.

\subsection{Method}

We initially determined the upper bounds on the guessing probabilities
$P_{\guess}(A_{1} | \rE)$ and $P_{\guess}(A_{1} B_{1} | \rE)$ numerically in
terms of the Mermin expectation value $\merminr$. It was quickly apparent
that the nonlinear parts of the bounds were consistently being attained with
anticommuting measurements. From there it was not difficult to guess the
optimal states and see that the numeric bounds seemed to coincide with the
(at this point, conjectured) analytic forms \eqref{eq:pa1_mermin} and
\eqref{eq:pa1b1_mermin} given in section~\ref{sec:results}. Experimenting a
little, we found that the bounds seemed to be attained respectively at the
NPA hierarchy levels $1 + \rAB + \rAC$ and $1 + \rAB + \rAC + \rBC$; this
told us that we should be able to find sum-of-squares decompositions out of
the operators at these levels for the tangents of the bounds.

We searched for sum-of-squares decompositions following a method similar to
\cite{ref:bp2015}. The idea is essentially to write the general form of a
candidate sum-of-squares decomposition in terms of unknown parameters, assert
that it should expand to the operator we want to show is positive
semidefinite, and then find parameters for which the assertion becomes true.

Using the tangents of the local-guessing-probability bound as an example, we
were searching for a solution to the problem
\begin{equation}
  \label{eq:candidate_sos}
  T - \sum_{i} P^{s\,2}_{i} = 0 \,,
\end{equation}
where $T$ is the target expansion \eqref{eq:a1_mermin_target}, for operators
$P^{\pm}_{i}$ of the form
\begin{equation}
  P^{s}_{i} = \sum_{j} c^{s}_{ij} R^{s}_{j} \,,
\end{equation}
where the $c^{s}_{ij}$s are unknown real-valued coefficients and the
$R^{s}_{j}$s form a basis of the space of linear combinations of the
operators at level $1 + \rAB + \rAC$ with the property
\begin{equation}
  R^{s}_{j} \ket{\Psi} = 0
\end{equation}
for the (conjectured) optimal measurements $A_{x}$, $B_{y}$, $C_{z}$ and
state $\ket{\Psi}$ described in subsection~\ref{sec:pa1_linearisation}. Such
a basis of $R^{s}_{j}$s is given by
Eqs.~\eqref{eq:a1_Rfirst}--\eqref{eq:a1_Rlast}.

We have applied some simplifications to the problem above, following
\cite{ref:bp2015}. In particular, writing
\begin{equation}
  \sum_{i} P^{\dagger}_{i} P_{i}
  = \sum_{jkrs} M^{rs}_{jk} \, R^{r\,\dagger}_{j} R^{s}_{k}
\end{equation}
with
\begin{equation}
  M^{rs}_{jk} = \sum_{i} c^{r\,*}_{j} \, c^{s}_{k}
\end{equation}
for the potentially more general problem with
\begin{equation}
  P_{i} = \sum_{js} c^{s}_{ij} R^{s}_{j} \,,
\end{equation}
we have used that it is not restrictive to assume that the coefficients
$c^{s}_{ij}$ are real-valued and that the symmetry of the target operator
\eqref{eq:a1_mermin_target} under the transformation $\tau$
\eqref{eq:a1_mermin_symmetry} can be used to block diagonalise the matrix of
elements $M^{rs}_{jk}$.

We also applied another simplification: one can choose to set
$c^{s}_{ij} = 0$ for (for instance) $i < j$ or $i > j$. This corresponds to
choosing a Cholesky factorisation of the matrix of elements
$M^{s}_{jk} = \sum_{i} c^{s}_{ij} c^{s}_{ik}$.

Expanding the candidate sum-of-squares decomposition on the left-hand side of
\eqref{eq:candidate_sos} and requiring operator-by-operator that the
left-hand side is zero translates to imposing a number of quadratic equality
constraints on the coefficients $c^{s}_{ij}$. We used a Python module
\texttt{divars.py}, which we have included with this article, together with
SymPy, to automate this procedure and help simplify the resulting
constraints. We then repeatedly searched numerically for solutions to the
constraints, guessing and gradually introducing constraints on the
coefficients (e.g., trying $c^{s}_{ij} = 0$ for some coefficient or imposing
that two coefficients are equal to each other) until the numeric search
seemed to consistently return the same solution. Solving the remaining
constraints by hand got us the sum-of-squares decomposition given in
subsection~\ref{sec:pa1_linearisation}.

\section{Attacks against device-independent secret sharing}
\label{sec:secret_sharing}

Aside from fundamental interest, a second more practical motivation to
conduct the previous analysis was to construct a device-independent
secret-sharing protocol based on the Mermin inequality. However, we found
obstacles to this idea which we describe in the following section.

\subsection{Overview}

Secret sharing is a cryptographic task in which a secret (e.g., a
cryptographic key) is distributed among two or more parties in such a way
that a specified minimum number of parties must work together in order to
reconstruct it. Hillery, Bu{\ifmmode \check{z}\else \v{z}\fi{}}ek, and
Berthiaume (HBB) \cite{ref:hbb1999} proposed a quantum version of secret
sharing, analogous to the concept of quantum key distribution, in which the
security of the protocol is guaranteed by quantum physics. In the three-party
scheme of \cite{ref:hbb1999}, Alice, Bob, and Charlie share a GHZ state
$\ket{\Psi} = \frac{1}{\sqrt{2}} \bigro{\ket{111} + \ket{222}}$ and choose
inputs $x, y, z \in \{1, 2\}$ and measure $A_{x}$, $B_{y}$, and $C_{z}$,
where $A_{1} = B_{1} = C_{1} = \sx$ and $A_{2} = B_{2} = C_{2} = \sy$. In all
cases, Bob's and Charlie's measurement outcomes individually are uncorrelated
with Alice's. However, if Alice, Bob, and Charlie all measure $\sx$, or any
one of them measures $\sx$ and the other two measure $\sy$, then Bob and
Charlie can together determine Alice's result from the product of their own
measurement results. Quantum secret-sharing protocols can also be devised for
more than three parties, but we will discuss explicitly only the three-party
version here.

The state and measurements, and resulting correlations, of this protocol are
precisely those that maximally violate the Mermin-Bell inequality. For
readers familiar with both, it may seem natural to ask whether the security
of the HBB scheme can be proved device independently, i.e., without assuming
that the participants' devices are necessarily measuring $\sx$ and
$\sy$. There have indeed been proposals to design a device-independent
secret-sharing protocol based on the GHZ-paradox or other correlations
arising from GHZ states \cite{ref:ag2012,ref:gz2017,ref:rm2017}. However, we
found that the HBB scheme is completely insecure from a device-independent
point of view. The reason is that the secret-sharing protocol is intended to
still work, securely, if either Bob or Charlie (but not both) are dishonest,
and this differs from the usual Bell scenario where all the parties
participating in the Bell test are trusted.

If (say) Charlie is dishonest, he could attack the protocol in the usual ways
considered in the security analyses of device-independent protocols
(particularly, he could prepare a different state than the GHZ state and/or
arrange for Alice's and Bob's devices to perform different measurements than
$\sx$ and $\sy$). Moreover, since Charlie is also involved in the parameter
estimation (e.g., the estimation of the Mermin expectation value), he could
also act in ways that don't respect the normal conditions of a Bell test:
\begin{enumerate}
  \item \label{enum:bob_basis_attack} Charlie could wait until Bob declares
  which basis $y$ he measured in before declaring his own input $z$ and
  output $c$, and could perform different measurements on his system
  depending on which input $y$ Bob declared.
  \item \label{emum:input_correlated} Charlie could introduce correlations
  between his choice of input $z$ and the system prepared for the protocol,
  instead of choosing $z$ randomly and independently, for instance by
  performing a four-outcome measurement to determine both his input $z$ and
  output $c$, or by implementing a hidden-variable model in which the hidden
  variable $\lambda$ is correlated with $z$.
  \item \label{enum:attack_input} Charlie could perform a different
  measurement to attempt to guess Alice's outcome than he does in the
  parameter estimation rounds.
\end{enumerate}
The possibility of an attack combining \ref{enum:bob_basis_attack} and
\ref{enum:attack_input} is already known to be fatal for even the
device-dependent HBB scheme (i.e., Charlie can learn Alice's outcome, without
being detected, even assuming that Alice and Bob are measuring $\sx$ and
$\sy$). It and a possible remedy, in which Bob and Charlie are required to
declare their outputs before either are allowed to declare their inputs, is
discussed in~\cite{ref:kki1999}.

In the following we describe how a dishonest party could go about attacking
an HBB-type protocol, in either the quantum or no-signalling scenarios,
without needing to learn Bob's input. We have not attempted to be exhaustive
or general; we merely describe the simplest pathological cases that would
need to be ruled out, which already show that the situation is much worse for
secret sharing in the device-independent scenario.

\subsection{Hidden variable models}

Similarly to other device-independent cryptographic protocols, the simplest
way a dishonest Charlie could try to attack a secret-sharing protocol would
be to attempt to implement a deterministic hidden-variable model replicating
the observed correlations. This is possible if the probabilities
$P(abc | xyz)$ of the protocol can be expressed in the form
\begin{IEEEeqnarray}{l}
  \label{eq:secret_sharing_hv}
  P(abc | xyz) \IEEEnonumber \\
  \qquad = \sum_{\lambda} p_{\lambda | z}
  P_{\rA}(a | x; \lambda) P_{\rB}(b | y; \lambda) P_{\rC}(c | z; \lambda)
  \,. \IEEEeqnarraynumspace
\end{IEEEeqnarray}
Note that, in this case, there is no reason for Charlie to arrange for the
hidden variable $\lambda$ and his own input $z$ to be uncorrelated. (In the
language of Bell locality, the so-called ``free will'' assumption is not
justified.) We reflect this in \eqref{eq:secret_sharing_hv} by allowing the
probability distribution $p_{\lambda | z}$ to depend arbitrarily on
$z$. Eq.~\eqref{eq:secret_sharing_hv} thus does not have the form of a local
hidden-variable model of the kind normally considered in Bell-type theorems,
and it is not sufficient for the probabilities $P(abc | xyz)$ to violate a
Bell inequality, such as the Mermin inequality, in order to rule out a local
hidden-variable model of the form above.

It is easy to show that the existence of a decomposition of the form
\eqref{eq:secret_sharing_hv} is equivalent to the existence of a local
hidden-variable model of the form
\begin{IEEEeqnarray}{l}
  P(ab | xy; cz) \IEEEnonumber \\
  \qquad = \sum_{\lambda} p'_{\lambda | c z}
  P^{(cz)}_{\rA}(a | x; \lambda) P^{(cz)}_{\rB}(b | y; \lambda)
  \IEEEeqnarraynumspace
\end{IEEEeqnarray}
for each of the probability distributions $P(ab | xy; cz)$ conditioned on
Charlie's different possible outputs and inputs $c$ and $z$. This gives a
bare minimum condition in order for there to be any hope that a
device-independent secret-sharing scheme might be secure: \emph{at least} one
of the conditional distributions $P(ab | xy; cz)$ (for some $c$ and $z$) must
be nonlocal. This condition is not met for the GHZ correlations that the HBB
protocol is based on: in that case all of the conditional distributions
$P(ab | xy; cz)$ exhibit perfect correlation or no correlation at all
depending on the inputs, and admit trivial local hidden-variable models. This
makes it clear that secret sharing cannot be done securely and device
independently using only the correlations of the GHZ paradox.

\subsection{No-signalling attacks}

Security analyses of device-independent protocols are sometimes undertaken
using only the no-signalling constraints, since this is typically much
simpler, though typically at the cost of significantly worse tolerance to
noise. We are aware of at least two proposals \cite{ref:ag2012,ref:gz2017} to
design device-independent secret-sharing protocols using GHZ states (but not
necessarily the GHZ-paradox correlations) using only no-signalling
constraints. In this case, the situation is significantly worse, since in the
no-signalling scenario, a dishonest Charlie could implement arbitrary
steering. More precisely, suppose Charlie wishes to produce the no-signalling
distribution $P(abc|xyz)$ in the parameter estimation rounds. If the marginal
distribution $P(ab | xy) = \sum_{c} P(abc | xyz)$ can be expressed as a
convex sum
\begin{equation}
  P(ab | xy) = \sum_{\lambda} p_{\lambda} P^{(\lambda)}(ab | xy)
\end{equation}
of no-signalling distributions $P^{(\lambda)}(ab|xy)$ then Charlie could
prepare the extended distribution
\begin{equation}
  \label{eq:nosig_attack}
  P'(abc | xyz) = \begin{cases}
    P(abc | xyz) &\text{if } z \neq \bot \\
    p_{c} P^{(c)}(ab | xy) &\text{if } z = \bot
  \end{cases}
\end{equation}
where $\bot$ is an additional input that Charlie can use in the secret bit
generation rounds, when he is not asked to publicly disclose his input and
outcome. It is easy to verify that the extended distribution
\eqref{eq:nosig_attack} still satisfies the no-signalling constraints.

The above observation means that, in the no-signalling scenario, the security
or insecurity of a device-independent secret-sharing protocol against a
dishonest Charlie is determined entirely by the marginal distribution
$P(ab|xy)$ between Alice and Bob. If this marginal distribution is in the
local polytope then the protocol is completely insecure against no-signalling
attacks. A special case worth remarking is that no device-independent
secret-sharing protocol based on the GHZ state can be proved secure using
only the no-signalling conditions: the marginals of the GHZ state are all
separable and the marginal probability distributions will always be in the
local polytope, regardless of what or how many measurements are performed by
the parties.

\subsection{Outlook}

To summarise: we have pointed out that a device-independent version of the
HBB protocol would be completely insecure against a dishonest party, and that
any protocol for which the marginal probability distributions are in the
local polytope (for example, any protocol using a GHZ state) cannot be proved
secure using only the no-signalling constraints. This does not rule out that
a device-independent secret-sharing protocol could be designed, for instance
based on different correlations and/or using stronger constraints than only
the no-signalling conditions in the security proof. However, one should
consider the following points:
\begin{itemize}
  \item It is already known that if one can do quantum key distribution then
  one can do secret sharing. For instance, Alice could do device-independent
  key distribution separately with Bob and Charlie and xor the two keys. More
  generally, secret sharing can be done securely using classical protocols if
  the parties can do one-time-pad encryption, which happens to be precisely
  what key distribution schemes are intended to generate cryptographic keys
  for.
  \item As with key distribution, or any secure protocol involving parties
  communicating remotely, the parties would need to authenticate
  themselves. This is normally done in key distribution using classical
  authentication schemes which require preshared keys; part of the generated
  key can then be used to do the authentication the next time. Consequently,
  it seems to us that one would need to be able to do key distribution anyway
  in order to do secret sharing, if only to generate the authentication keys
  needed after the first use of the protocol.
\end{itemize}
Given these issues, the usefulness of a device-independent secret-sharing
protocol that does not reduce to a direct application of device-independent
quantum key distribution is unclear to us.

\section{Conclusion}

We considered the Mermin-Bell experiment with three parties and we identified
and proved tight upper bounds on the guessing probabilities associated with
the measurement outcomes of one and two of the parties. The results are
fundamental tradeoffs between the amount of intrinsic randomness and
nonlocality, as measured by the violation of the Mermin inequality, imposed
by the structure of quantum physics. The linearisations in
section~\ref{sec:tangents} can also be read as inequalities identifying parts
of the boundary of the set of quantum correlations. The results reveal that
part of the boundary of the quantum set is flat, a characteristic that has
previously been remarked upon in \cite{ref:rm2017,ref:gk2018}.

It may be interesting to study how our results generalise to Bell experiments
involving more parties. We guessed one possible generalisation of the upper
bound \eqref{eq:pa1b1_mermin} for $P_{\guess}(A_{1} B_{1} | \rE)$ to $n$
parties, which can be found in appendix~\ref{sec:pguess_n_parties}. We did
not attempt to prove it, though we tested the cases for $n = 4$ and $5$
parties numerically.

While we are not aware of an obvious practical application of our results, we
believe there is some merit to finding the analytic form of randomness vs.\
nonlocality tradeoffs more generally where it could be feasible to do so,
particularly where the result might be used in the security proof of a
device-independent protocol. From this point of view, our work has explored
the feasibility of searching for sum-of-squares decompositions for problems
somewhat larger than was considered in \cite{ref:bp2015}. The cases where the
method is likely to work are probably those where the problem is ``simple''
in some same key respects as the problems we studied. In particular: it was
reasonably easy for us to guess the upper bounds and the states and
measurements that attained them, we found that the optimal solution was
attained at a level of the hierarchy that was not prohibitively high, and
symmetries of the problem allowed us to reduce the number of variables in the
searches for sum-of-squares decompositions.

\section*{Acknowledgements}

E.W. thanks Cedric Bamps and Stefano Pironio for helpful discussions
concerning sum-of-squares decompositions. This work was supported by the
Spanish MINECO (QIBEQI FIS2016-80773-P and Severo Ochoa grant SEV-2015-0522),
the Generalitat de Catalunya (CERCA Program and SGR 1381), the Fundaci\'o{}
Privada Cellex, the AXA Chair in Quantum Information Science, and the ERC CoG
QITBOX. B.B. acknowledges support from the Secretaria d'Universitats i
Recerca del Departament d'Economia i Coneixement de la Generalitat de
Catalunya and the European Social Fund (FEDER).

\bibliography{mermin_rng}

\appendix

\section{No-signalling bounds}
\label{sec:nosignalling_bounds}

The main text gave tight bounds on the guessing probability assuming all the
measurements are performed on a quantum system. The tightest bound that can
be derived for the local guessing probability using only the no-signalling
constraints is
\begin{equation}
  \label{eq:pa1_nosig}
  P_{\guess}(A_{1} | \rE)
  \leq \frac{3}{2} - \frac{1}{8} \abs{\merminr}
  - \frac{1}{8} \abs{\mermini} \,.
\end{equation}
For the two-party guessing probability there are two distinct tight bounds,
\begin{equation}
  \label{eq:pa1b1_nosig1}
  P_{\guess}(A_{1} B_{1} | \rE) \leq \frac{3}{2}
  - \frac{1}{4} \abs{\merminr}
\end{equation}
and
\begin{equation}
  \label{eq:pa1b1_nosig2}
  P_{\guess}(A_{1} B_{1} | \rE) \leq \frac{7}{4} 
  - \frac{1}{4} \abs{\merminr} - \frac{1}{8} \abs{\mermini} \,,
\end{equation}
as well as the same bounds with $\merminr$ and $\mermini$ swapped. Finally,
there are three bounds for the global guessing probability,
\begin{IEEEeqnarray}{rCl}
  \label{eq:pa1b1c1_nosig1}
  P_{\guess}(A_{1} B_{1} C_{1} | \rE)
  &\leq& \frac{3}{2} - \frac{1}{4} \abs{\merminr} \,, \\
  \label{eq:pa1b1c1_nosig2}
  P_{\guess}(A_{1} B_{1} C_{1} | \rE) &\leq& \frac{7}{4} - \frac{1}{4}
  \abs{\merminr}
  - \frac{1}{8} \abs{\mermini} \,, \\
  \label{eq:pa1b1c1_nosig3}
  P_{\guess}(A_{1} B_{1} C_{1} | \rE) &\leq& \frac{7}{4} - \frac{1}{16}
  \abs{\merminr} - \frac{5}{16} \abs{\mermini} \,. \IEEEeqnarraynumspace
\end{IEEEeqnarray}
The same upper bounds hold for $P_{\guess}(A_{1} B_{2} C_{2} | \rE)$,
$P_{\guess}(A_{2} B_{1} C_{2} | \rE)$, and
$P_{\guess}(A_{2} B_{2} C_{1} | \rE)$. The upper bounds for
$P_{\guess}(A_{1} B_{1} C_{2} | \rE)$, $P_{\guess}(A_{1} B_{2} C_{1} | \rE)$,
$P_{\guess}(A_{2} B_{1} C_{1} | \rE)$, and
$P_{\guess}(A_{2} B_{2} C_{2} | \rE)$ are the same except with $\merminr$ and
$\mermini$ swapped.

Following an approach similar to \cite{ref:spm2013}, the
local-guessing-probability bound \eqref{eq:pa1_nosig} is implied by the eight
inequalities
\begin{IEEEeqnarray}{rCl}
  \label{eq:a1_b1c1_nosig}
  1 - \avg{A_{1}} + \avg{B_{1} C_{1}} - \avg{A_{1} B_{1} C_{1}}
  &\geq& 0 \,, \IEEEeqnarraynumspace \\
  1 - \avg{A_{1}} + \avg{B_{1} C_{2}} - \avg{A_{1} B_{1} C_{2}}
  &\geq& 0 \,, \\
  1 - \avg{A_{1}} + \avg{B_{2} C_{1}} - \avg{A_{1} B_{2} C_{1}}
  &\geq& 0 \,, \\
  1 - \avg{A_{1}} - \avg{B_{2} C_{2}} + \avg{A_{1} B_{2} C_{2}}
  &\geq& 0 \,, \\
  1 + \avg{A_{2}} - \avg{B_{1} C_{1}} - \avg{A_{2} B_{1} C_{1}}
  &\geq& 0 \,, \\
  1 - \avg{A_{2}} - \avg{B_{1} C_{2}} + \avg{A_{2} B_{1} C_{2}}
  &\geq& 0 \,, \\
  1 - \avg{A_{2}} - \avg{B_{2} C_{1}} + \avg{A_{2} B_{2} C_{1}}
  &\geq& 0 \,, \\
  \label{eq:a2_b2c2_nosig}
  1 + \avg{A_{2}} + \avg{B_{2} C_{2}} + \avg{A_{2} B_{2} C_{2}} &\geq& 0 \,.
\end{IEEEeqnarray}
Each of these is in turn implied by two positivity constraints. For example,
\eqref{eq:a1_b1c1_nosig} is just stating that
\begin{equation}
  4 P({-}{+}{+}|111) + 4 P({-}{-}{-}|111) \geq 0 \,.
\end{equation}
The inequalities \eqref{eq:a1_b1c1_nosig} to \eqref{eq:a2_b2c2_nosig} sum to
\begin{equation}
  8 - 4 \avg{A_{1}} - \merminr - \mermini \geq 0
\end{equation}
which, together with symmetries of the problem, implies \eqref{eq:pa1_nosig}.

The upper bound $P_{\guess}(A_{1}B_{1} | \rE) \leq 3/2 - \merminr/4$ is
similarly implied by the five inequalities
\begin{IEEEeqnarray}{rCl}
  1 + \avg{C_{1}} - \avg{A_{1} B_{1}} - \avg{A_{1} B_{1} C_{1}}
  &\geq& 0 \,, \IEEEeqnarraynumspace \\
  1 - \avg{A_{1}} - \avg{B_{2} C_{2}} + \avg{A_{1} B_{2} C_{2}}
  &\geq& 0 \,, \\
  1 - \avg{B_{1}} - \avg{A_{2} C_{2}} + \avg{A_{2} B_{1} C_{2}}
  &\geq& 0 \,, \\
  1 - \avg{C_{1}} - \avg{A_{2} B_{2}} + \avg{A_{2} B_{2} C_{1}}
  &\geq& 0 \,, \\
  1 + \avg{A_{2} B_{2}} + \avg{A_{2} C_{2}} + \avg{B_{2} C_{2}} &\geq& 0
\end{IEEEeqnarray}
(the last of these is just stating that
\begin{equation}
  4 P({+}{+}{+}|222) + 4 P({-}{-}{-}|222) \geq 0 \,),
\end{equation}
which sum to
\begin{equation}
  6 - 4 P_{\rAB}({+}{+}|11) - \merminr \geq 0 \,.
\end{equation}
The second upper bound \eqref{eq:pa1b1_nosig2} for
$P_{\guess}(A_{1} B_{1} \rE)$ is implied by the inequalities
\begin{IEEEeqnarray}{rCl}
  2 + 2 \avg{C_{1}} - 2 \avg{A_{1} B_{1}} - 2 \avg{A_{1} B_{1} C_{1}}
  &\geq& 0 \,, \IEEEeqnarraynumspace \\
  1 - \avg{A_{1}} + \avg{B_{1} C_{2}} - \avg{A_{1} B_{1} C_{2}}
  &\geq& 0 \,, \\
  1 - \avg{C_{1}} + \avg{A_{1} B_{2}} - \avg{A_{1} B_{2} C_{1}}
  &\geq& 0 \,, \\
  1 - \avg{B_{1}} + \avg{A_{2} C_{1}} - \avg{A_{2} B_{1} C_{1}}
  &\geq& 0 \,, \\
  \label{eq:nosig_example}
  2 - \avg{A_{1}} - \avg{C_{2}} - \avg{A_{1} B_{2}} && \IEEEnonumber \\
  -\> \avg{B_{2} C_{2}} + 2 \avg{A_{1} B_{2} C_{2}} &\geq& 0 \,, \\
  2 - \avg{A_{2}} - \avg{B_{1}} - \avg{A_{2} C_{2}} && \IEEEnonumber \\
  -\> \avg{B_{1} C_{2}} + 2 \avg{A_{2} B_{1} C_{2}} &\geq& 0 \,, \\
  2 - \avg{B_{2}} - \avg{C_{1}} - \avg{A_{2} B_{2}} && \IEEEnonumber \\
  -\> \avg{A_{2} C_{1}} + 2 \avg{A_{2} B_{2} C_{1}} &\geq& 0 \,, \\
  1 + \avg{A_{2}} + \avg{B_{2}} + \avg{C_{2}} + \avg{A_{2} B_{2}}
  && \IEEEnonumber \\
  +\> \avg{A_{2} C_{2}} + \avg{B_{2} C_{2}} + \avg{A_{2} B_{2} C_{2}} &\geq&
  0 \,,
\end{IEEEeqnarray}
which sum to
\begin{equation}
  14 - 8 P_{\rAB}({+}{+}|11) - 2 \merminr - \mermini \geq 0 \,.
\end{equation}
Each of the eight inequalities above can be obtained from up to three
positivity constraints. For instance, the left-hand side of
\eqref{eq:nosig_example} is equal to
\begin{IEEEeqnarray}{rl}
  4 P({+}{-}{-} | 122) + 8 P({-}{+}{-} | 122) & \IEEEnonumber \\
  +\> 4 P({-}{-}{+} | 122) & \,.
\end{IEEEeqnarray}

The first two upper bounds \eqref{eq:pa1b1c1_nosig1} and
\eqref{eq:pa1b1c1_nosig2} on the three-outcome guessing probability
$P_{\guess}(A_{1} B_{1} C_{1} | \rE)$ are implied by \eqref{eq:pa1b1_nosig1}
and \eqref{eq:pa1b1_nosig2}. Using symmetries of the problem, the remaining
inequality \eqref{eq:pa1b1c1_nosig3} reduces to showing that
\begin{multline}
  \max \bigro{P({+}{+}{+} | 111), P({-}{-}{-} | 222)} \\
  \leq \frac{7}{4} - \frac{1}{16} M - \frac{1}{16} M' \,.
\end{multline}
One can readily verify that
\begin{IEEEeqnarray*}{rCl}
  \IEEEeqnarraymulticol{3}{l}{\tfrac{7}{4}
    - P({+}{+}{+} | 111) - \tfrac{1}{16} \merminr - \tfrac{5}{16} \mermini} \\
  \qquad &=&
  \tfrac{1}{4} P({+}{+}{-} | 111) + \tfrac{1}{4} P({+}{-}{+} | 111) \\
  &&+\> \tfrac{1}{4} P({-}{+}{+} | 111) + \tfrac{3}{4} P({-}{-}{-} | 111)
  \IEEEeqnarraynumspace \\
  &&+\> P({+}{-}{+} | 112) + P({-}{+}{+} | 112) \\
  &&+\> \tfrac{1}{2} P({-}{-}{-} | 112) \\
  &&+\> P({+}{+}{-} | 121) + P({-}{+}{+} | 121) \\
  &&+\> \tfrac{1}{2} P({-}{-}{-} | 121) \\
  &&+\> P({+}{+}{-} | 211) + P({+}{-}{+} | 211) \\
  &&+\> \tfrac{1}{2} P({-}{-}{-} | 211) \\
  &&+\> \tfrac{1}{2} P({+}{-}{-} | 122) + \tfrac{1}{2} P({-}{+}{-} | 212) \\
  &&+\> \tfrac{1}{2} P({-}{-}{+} | 221) \\
  &&+\> \tfrac{1}{4} P({+}{+}{+} | 222) + \tfrac{3}{4} P({+}{-}{-} | 222) \\
  &&+\> \tfrac{3}{4} P({-}{+}{-} | 222) + \tfrac{3}{4} P({-}{-}{+} | 222) \\
  &\geq& 0 \IEEEyesnumber
\end{IEEEeqnarray*}
and
\begin{IEEEeqnarray*}{rCl}
  \IEEEeqnarraymulticol{3}{l}{\tfrac{7}{4}
    - P({-}{-}{-} | 111) - \tfrac{1}{16} \merminr
    - \tfrac{5}{16} \mermini} \\
  \qquad &=&
  \tfrac{1}{2} P({+}{+}{+} | 111) \\
  &&+\> \tfrac{1}{4} P({+}{+}{-} | 112) + P({+}{-}{+} | 112)
  \IEEEeqnarraynumspace \\
  &&+\> P({-}{+}{+} | 112) + \tfrac{1}{4} P({-}{-}{-} | 112) \\
  &&+\> P({+}{+}{-} | 121) + \tfrac{1}{4} P({+}{-}{+} | 121) \\
  &&+\> P({-}{+}{+} | 121) + \tfrac{1}{4} P({-}{-}{-} | 121) \\
  &&+\> P({+}{+}{-} | 211) + P({+}{-}{+} | 211) \\
  &&+\> \tfrac{1}{4} P({-}{+}{+} | 211) + \tfrac{1}{4} P({-}{-}{-} | 211) \\
  &&+\> \tfrac{1}{4} P({+}{+}{+} | 122) + \tfrac{1}{4} P({+}{-}{-} | 122) \\
  &&+\> \tfrac{1}{4} P({+}{+}{+} | 212) + \tfrac{1}{4} P({-}{+}{-} | 212) \\
  &&+\> \tfrac{1}{4} P({+}{+}{+} | 221) + \tfrac{3}{4} P({-}{-}{+} | 221) \\
  &&+\> \tfrac{1}{2} P({-}{-}{-} | 221) \\
  &&+\> P({+}{-}{-} | 222) + P({-}{+}{-} | 222) \\
  &&+\> \tfrac{1}{2} P({-}{-}{+} | 222) \\
  &\geq& 0 \IEEEyesnumber
\end{IEEEeqnarray*}
under the no-signalling constraints.

The bounds given here are the tightest that can be derived given that there
are no-signalling distributions for which
\begin{IEEEeqnarray*}{rlrrrl}
  \bigro{P_{\guess}(A_{1} | \rE), \merminr, \mermini} \in \bigbr{
    & \bigro{&1,\; &  0,\; & \pm' 4 &} , \\
    & \bigro{&1,\; & \pm 4,\; & 0 &} , \\
    & \bigro{&\tfrac{1}{2},\; & \pm 4,\; & \pm' 4 &}} \,,
  \IEEEeqnarraynumspace \IEEEyesnumber
\end{IEEEeqnarray*}
\begin{IEEEeqnarray*}{rlrrrl}
  \bigro{P_{\guess}(A_{1} B_{1} | \rE), \merminr, \mermini} \in \bigbr{
    & \bigro{&1,\; & \pm 2,\; & \pm' 2 &} , \\
    & \bigro{&\tfrac{1}{2},\; & \pm 2,\; & \pm' 4 &} , \\
    & \bigro{&\tfrac{1}{2},\; & \pm 4,\; & \pm' 2 &} , \\
    & \bigro{&\tfrac{1}{4},\; & \pm 4,\; & \pm' 4 &}} \,,
  \IEEEeqnarraynumspace \IEEEyesnumber
\end{IEEEeqnarray*}
and
\begin{IEEEeqnarray*}{rlrrrl}
  \bigro{P_{\guess}(A_{1} B_{1} C_{1} | \rE), \merminr, \mermini} \in \bigbr{
    & \bigro{&1,\; & \pm 2,\; & \pm' 2 &} , \\*
    & \bigro{&\tfrac{1}{2},\; & \pm 4,\; & \pm' 2 &} , \\*
    & \bigro{&\tfrac{1}{2},\; & 0,\; & \pm' 4 &} , \\*
    & \bigro{&\tfrac{1}{4},\; & \pm 4,\; & \pm' 4 &}} \,,
  \IEEEeqnarraynumspace \IEEEyesnumber
\end{IEEEeqnarray*}
The only case that might not be immediately obvious is that there are
no-signalling distributions for which simultaneously
$P_{\guess}(A_{1} B_{1} | \rE) = 1/2$, $\merminr = \pm 4$, and
$\mermini = \pm' 2$; these can be attained with vertices of class~34
according to the classification used in table~1 of \cite{ref:pbs2011}.

\section{Possible bound for \lowercase{$n > 3$} parties}
\label{sec:pguess_n_parties}

In section~\ref{sec:pa1b1_linearisation} we showed that the upper bound
\eqref{eq:pa1b1_mermin} on the two-party guessing probability
$P_{\guess}(A_{1} B_{1} | \rE)$ is tight and the nonlinear part $M \geq 3$
can be attained if the parties measure $\sx$ and $\sy$ on a state of the form
\begin{equation}
  \ket{\Psi} = \lambda \ket{+{+}+}
  + \mu \bigro{\ket{+{-}-} + \ket{-{+}-} + \ket{-{-}+}} \,.
\end{equation}
We mention a possible extension here for the $n$-partite Mermin correlator
\begin{equation}
  \merminr_{n} = \re \biggsq{
    \Bavg{\prod_{p=1}^{n} \bigro{A^{(p)}_{1} + i A^{(p)}_{2}}}} \,,
\end{equation}
where $A^{(p)}_{x}$ are the $p$th party's measurement operators, whose local
and quantum bounds are respectively \cite{ref:m1990}
\begin{equation}
  L_{n} = \begin{cases}
    2^{(n - 1)/2} &\text{if } n \text{ odd} \\
    2^{n/2} &\text{if } n \text{ even}
  \end{cases}
\end{equation}
and
\begin{equation}
  Q_{n} = 2^{n-1}
\end{equation}
(although the local bound $M_{n} \leq L_{n}$ is a facet of the local polytope
only for odd $n$).

The obvious generalisation of the strategy of
section~\ref{sec:pa1b1_linearisation} is for the $n$ parties to measure
\begin{IEEEeqnarray}{c+c}
  A^{(p)}_{1} = \sx \,, & A^{(p)}_{2} = \sy
\end{IEEEeqnarray}
on an $n$-partite state of the form
\begin{equation}
  \ket{\Psi} = \lambda \ket{+}^{\otimes n}
  + \mu \sum_{\vect{s} \in \mathcal{S}} \ket{\vect{s}} \,,
\end{equation}
where $\mathcal{S} \subset \{+, -\}^{\times n}$ is the subset of all vectors
of $n$ signs with a nonzero even number of minuses. The state is normalised
if
\begin{equation}
  \lambda^{2} + (Q_{n} - 1) \mu^{2} = 1 \,.
\end{equation}
In terms of $\lambda$ and $\mu$, the probability that the first $n - 1$
parties (or all $n$ of them, for that matter) obtain the result `$+$' if they
measure $\sx$ is
\begin{equation}
  P_{\vect{\rA}}(\vect{+} | \vect{1}) = \lambda^{2}
\end{equation}
and the Mermin expectation value is
\begin{equation}
  \merminr_{n} = \bigro{\lambda + (Q_{n} - 1) \mu}^{2} \,.
\end{equation}
Relating $P_{\vect{\rA}}(\vect{+} | \vect{1}) = \lambda^{2}$ to $M_{n}$
yields the dependence $P_{\vect{\rA}}(\vect{1} | \vect{1}) = P_{n}(M_{n})$,
where
\begin{IEEEeqnarray}{rCl}
  P_{n}(\merminr_{n})
  &=& 1 - \frac{1}{Q_{n}} - \frac{Q_{n} - 2}{Q\du{n}{2}} \merminr_{n}
   \IEEEnonumber \\
   &&+\> 2 \frac{\sqrt{Q_{n} - 1}}{Q\du{n}{2}}
   \sqrt{\merminr_{n} (Q_{n} - \merminr_{n})} \,. \IEEEeqnarraynumspace
\end{IEEEeqnarray}
By suitably mixing this strategy with a deterministic strategy with
$P_{\vect{\rA}}(\vect{1} | \vect{1}) = 1$ and $\merminr_{n} = L_{n}$, we
obtain a strategy for which the guessing probability and Mermin expectation
value are related by
\begin{equation}
  \label{eq:pA1_mermin}
  P_{\guess}(\vect{A}_{\vect{1}} | \rE)
  = \begin{cases}
    P_{n}(\merminr_{n})
    & \text{if } \merminr_{n} \geq \merminr_{n}^{\text{th}} \\
    \Gamma_{n}(\merminr_{n})
    & \text{if } \merminr_{n} \leq \merminr_{n}^{\text{th}}
  \end{cases} \,,
\end{equation}
where
\begin{equation}
  \Gamma_{n}(\merminr_{n})
  = \frac{L_{n} (Q_{n} - 1) - (L_{n} - 1) \merminr_{n}}{
    L_{n} (Q_{n} - L_{n})} \,.
\end{equation}
The threshold $\merminr_{n}^{\text{th}}$ in \eqref{eq:pA1_mermin} is the
point where the linear interpolation $\Gamma_{n}(\merminr_{n})$ coincides
with the curve $P_{n}(\merminr_{n})$ and their derivatives are the same. This
occurs at
\begin{equation}
  \label{eq:Mth_def}
  \merminr_{n}^{\text{th}}
  = \frac{L\du{n}{2} (Q_{n} - 1)}{L\du{n}{2} - 2 L_{n} + Q_{n}} \,,
\end{equation}
at which point
\begin{equation}
  P_{\guess}(\vect{A}_{\vect{1}} | \rE)
  = \frac{Q_{n} - 1}{L\du{n}{2} - 2 L_{n} + Q_{n}} \,.
\end{equation}
For odd $n$, we remark that $L_{n} = \sqrt{Q_{n}}$ and in that case
\eqref{eq:Mth_def} reduces to the average
$\merminr_{n}^{\text{th}} = (L_{n} + Q_{n}) / 2$ of the local and quantum
bounds.

The strategy we have described here shows that the upper bound on the
guessing probability cannot be better than \eqref{eq:pA1_mermin}. For $n = 4$
and $5$ parties, some numerical tests we carried out seemed to support that
the upper bound on the guessing probability coincides with
\eqref{eq:pA1_mermin}, although we did not attempt to prove this.

\end{document}